\documentclass{article}
\usepackage{amsmath}
\usepackage{cite}
\usepackage{graphicx}
\usepackage{dcolumn}

\begin{document}

\date{}
\title{On the variational treatment of a class of double-well oscillators}
\author{Francisco M. Fern\'{a}ndez\thanks{fernande@quimica.unlp.edu.ar} \\
INIFTA, DQT, Sucursal 4, C. C. 16, \\
1900 La Plata, Argentina \\
Javier Garcia \\
Instituto de F\'isica La Plata, \\
Diagonal 113 entre 63 y 64, 1900 La Plata, Argentina}
\maketitle

\begin{abstract}
We compare the well known Rayleigh-Ritz variational method (RRVM) with a
recently proposed approach based on supersymmetric quantum mechanics and the
Gram-Schmidt orthogonalization method (SSQMGS). We apply both procedures to
a particular class of double-well harmonic oscillators that had been
conveniently chosen for the application of the latter approach. The RRVM
eigenvalues converge smoothly from above providing much more accurate
results with less computational effort. Present results show that the
unproved SSQMGS upper bounds do not hold.
\end{abstract}

\section{Introduction}

\label{sec:intro}

There has been some interest in the calculation of the eigenvalues and
eigenfunctions of rather simple one-dimensional Hamiltonians with
double-well potentials because they are supposed to be useful for the
calculation of the probability density for the one-dimensional Fokker-Planck
equation\cite{BFR10,S16}. In a paper appeared recently in this journal
Batael and Drigo Filho\cite{BF23} proposed a variational method that is
supposed to yield upper bounds to all the eigenvalues of the Hamiltonian.
They constructed the variational wavefunctions by means of supersymmetric
quantum mechanics (SSQM) and the Gram-Schmidt (GM) orthogonalization method
but did not provide a plausible proof for those bounds. From now, on we will
refer to this approach as SSQMGS.

The well known Rayleigh-Ritz variational method (RRVM), discussed in most
textbooks on quantum chemistry\cite{P68,SO96}, is known to provide upper
bounds to all the eigenvalues of a given Hamiltonian operator\cite{M33} (see
also a recent simpler proof of the RRVM upper bounds\cite{F20}). In our
opinion, it is interesting to compare the SSQMGS and the RRVM because they
apparently exhibit somewhat similar features.

In section~\ref{sec:models_methods} we discuss the one-dimensional
quantum-mechanical models and the approximate method used for obtaining
their eigenvalues. In section~\ref{sec:results} we compare and discuss the
eigenvalues provided by SSQMGS and RRVM.

\section{Models and methods}

\label{sec:models_methods}

Batael and Drigo Filho\cite{BF23} obtained some eigenvalues of a particular
class of simple quantum-mechanical models of the form
\begin{equation}
H=-\frac{d^{2}}{dx^{2}}+V(x),\;V(x)=\sum_{j=0}^{2K-1}A_{2j}x^{2j},\;K=2,3,
\label{eq:H_general}
\end{equation}
that are amenable for the application of SSQM. We can provide an alternative
sound reason for this choice without resorting to SSQM.

From a square-integrable exponential function
\begin{equation}
\psi _{0}(x)=e^{-F(x)},  \label{eq:psi_0}
\end{equation}
we can obtain a reference potential $V_{0}(x)$ as follows
\begin{equation}
\frac{\psi _{0}^{\prime \prime }}{\psi _{0}}=F^{\prime 2}-F^{\prime \prime
}=V_{0}-E_{0},\;F(x)=\sum_{j=0}^{K}F_{j}x^{2j},\;F_{K}>0.  \label{eq:V_0}
\end{equation}
For some particular values of the coefficients $A_{j}$ we can choose $F_{j}$
so that $V_{0}(x)=V(x)$. In this case both $\psi _{0}(x)$ and $E_{0}$ are
exact and the SSQMGS is expected to yield the most accurate results.

For example, when $K=2$ the requirement for $V_{0}(x)=V(x)$ is that the
coefficients $A_{j}$ satisfy
\begin{equation}
4a_{2}a_{6}-a_{4}^{2}+12a_{6}^{3/2}=0,  \label{eq:rest_cond_x6}
\end{equation}
while for $K=3$ we have two restrictive conditions
\begin{eqnarray}
4a_{10}a_{6}a_{8}-40a_{10}^{5/2}-8a_{10}^{2}a_{4}-a_{8}^{3} &=&0,  \nonumber
\\
16a_{10}^{2}a_{6}^{2}-64a_{10}^{3}a_{2}-96a_{8}a_{10}^{5/2}-8a_{10}a_{6}a_{8}^{2}+a_{8}^{4} &=&0.
\label{eq:rest_cond_x10}
\end{eqnarray}
The role of $A_{0}$ is irrelevant because it is just a shift of $E_{0}$.

Since the potential $V(x)$ is parity invariant, then the eigenfunctions of $H
$ have definite parity (they are either even or odd). This fact enables us
to apply the approximate methods to each symmetry thus reducing considerably
the computation time.

The RRVM is based on trial functions of the form
\begin{equation}
\varphi ^{[N]}=\sum_{j=0}^{N-1}c_{j}f_{j},  \label{eq:RRVM_trial}
\end{equation}
where $B=\left\{ f_{0},f_{1},\ldots \right\} $ is a complete set of basis
functions. The variational principle leads to a secular equation of the form
\begin{equation}
\left( \mathbf{H}-E\mathbf{S}\right) \mathbf{c},  \label{eq:RRVM_sec_eq}
\end{equation}
where the $N\times N$ matrices $\mathbf{H}$ and $\mathbf{S}$ have elements $%
H_{ij}=\left\langle f_{i}\right| H\left| f_{j}\right\rangle $ and $%
S_{ij}=\left\langle f_{i}\right. \left| f_{j}\right\rangle $, respectively,
and $\mathbf{c}$ is a column vector of the expansion coefficients $c_{j}$%
\cite{P68,SO96}. The approximate eigenvalues $E_{i}^{[N]}$, $i=0,1,\ldots
,N-1$, are roots of the secular determinant $\left| \mathbf{H}-E\mathbf{S}%
\right| $ that yields the characteristic polynomial $p(E)$. It can be proved
that $E_{n}^{[N-1]}>E_{n}^{[N]}>E_{n}$, where $E_{n}$ is an exact eigenvalue
of $H$\cite{M33,F20}. For each $E_{n}^{[N]}$ we obtain $\varphi _{n}^{[N]}$
and it can be proved that $\left\langle \varphi _{i}^{[N]}\right. \left|
\varphi _{j}^{[N]}\right\rangle =0$ if $E_{i}^{[N]}\neq E_{j}^{[N]}$.
Obviously, in the case of present one-dimensional toy models there are no
degenerate states and $\left\langle \varphi _{i}^{[N]}\right. \left| \varphi
_{j}^{[N]}\right\rangle =0$ if $i\neq j$. In particular, when the basis set $%
B$ is orthonormal then $\mathbf{S}=\mathbf{I}$ (the $N\times N$ identity
matrix).

It follows from $\left\langle \varphi _{i}^{[N]}\right| H\left| \varphi
_{j}^{[N]}\right\rangle =\left\langle \varphi _{i}^{[N]}\right. \left|
\varphi _{j}^{[N]}\right\rangle E_{i}^{[N]}$\cite{F20} that
\begin{equation}
E_{i}^{[N]}=\frac{\left\langle \varphi _{i}^{[N]}\right| H\left| \varphi
_{i}^{[N]}\right\rangle }{\left\langle \varphi _{i}^{[N]}\right. \left|
\varphi _{i}^{[N]}\right\rangle }>E_{i},  \label{eq:upper_bound}
\end{equation}
that resembles the SSQMGS expression (equation (8)) reported by
Batael and Drigo Filho\cite{BF23} without proof. More precisely,
one can prove rigorously that the bounds proposed by these authors
apply to the ground state and first-excited state that have the
smallest energy for each symmetry (provided, of course, that the
trial functions have the appropriate symmetry). However, as far as
we know, there is no proof for the remaining states (as in the
case of the RRVM\cite{M33,F20}). The outcome of upper bounds and
orthogonal approximate wavefunctions make the RRVM and the SSQMGS
look similar, with the difference that in the latter case the
upper bounds have not been rigorously proved, except in the two
cases just mentioned.

When the potential is parity invariant, we can apply the RRVM to each kind
of symmetry thus reducing the dimension of the matrices involved in the
calculation.

The simplest basis set is given by the eigenfunctions of the Harmonic
oscillator
\begin{equation}
H_{HO}=-\frac{d^{2}}{dx^{2}}+\omega ^{2}x^{2}.  \label{eq:H_HO}
\end{equation}
Although the asymptotic behaviour of the eigenfunctions of $H_{HO}$ is quite
different from that of the problems discussed here, such eigenfunctions
exhibit two advantages. First, we already know that this orthonormal basis
set is complete and, second, the matrix elements $H_{ij}$ can be calculated
exactly without difficulty.

In principle, we can resort to a set of basis functions with suitable
asymptotic behaviour\cite{FG14} but it is not necessary for present
discussion.

There are many ways of obtaining a suitable value of $\omega $; here we
arbitrarily resort to the condition
\begin{equation}
\frac{d}{d\omega }\sum_{j=0}^{M}H_{jj}(\omega )=0,\;M\leq N\text{.}
\label{eq:omega_cond}
\end{equation}

\section{Results and discussion}

\label{sec:results}

We first consider the example with $K=2$, $A_{0}=1$, $A_{2}=A_{4}=-2$ and $%
A_{6}=1$ that allows an exact ground state $\psi _{0}$ with $E_{0}=0$. The
ansatz used by Batael and Drigo Filho\cite{BF23} is a curious linear
combination of functions with no definite parity which is not convenient for
a parity-invariant Hamiltonian operator. They chose the nonlinear parameter $%
c_{0}=0$ that leads to the exact ground-state eigenfunction $\psi _{0}$ (of
even parity) but it is not clear why they kept the nonlinear parameter $%
c_{n} $ in the exponential factors of the other trial functions $\psi _{n}$.
Although Batael and Drigo Filho mentioned the advantage of the separate
treatment of even and odd states, their ans\"{a}tze do not reflect this
fact. Another curious feature of their approach is the choice of Legendre
Polynomials that are known to be orthogonal in the interval $[-1,1]$ when in
the present case the variable interval is $(-\infty ,\infty )$.

Tables \ref{tab:model1e} and \ref{tab:model1o} show the convergence of the
lowest RRVM eigenvalues towards results that are supposed to be accurate up
to the last digit. We estimated $\omega $ from equation (\ref{eq:omega_cond}%
) with $M=10$ and arbitrarily chose an integer value close to the
real root. Although the rate of convergence depends on $\omega $,
the choice of an optimal value of this adjustable parameter is not
that relevant. As expected, the RRVM eigenvalues converge from
above\cite{P68,SO96,M33,F20}. Note that Batael and Drigo
Filho\cite{BF23} did not report the eigenvalues $E_{n}$ of this
model but $\lambda _{n}=E_{n}/2$ as in reference\cite{BFR10}.
Although the RRVM requires about 25 basis functions for a
ten-digit accuracy, the calculation is extremely simple because it
only requires the diagonalization of matrices $\mathbf{H}$ with
elements $H_{ij}$ that can be obtained analytically. On the other
hand, the SSQMGS is considerably cumbersome because it requires
the numerical calculation of all the integrals and minimization of
the approximate energy that is a function of nonlinear parameters.
Besides, the accuracy of these results cannot be improved any
further.

In the second example we also have $K=2$, but since $A_{0}=0$, $A_{2}=-26$, $%
A_{4}=6$ and $A_{6}=1$ then there is no exact ground state. Tables
\ref {tab:model2e} and \ref{tab:model2o} show the the convergence
of the lowest RRVM eigenvalues towards results that are also
supposed to be accurate up to the last digit. We estimated $\omega
$ as in the previous example. In this case, we appreciate that the
SSQMGS eigenvalues $E_{4}$ and $E_{6}$ do not provide upper bounds
which suggests that the equation (8) of Batael and Drigo
Filho\cite{BF23} does not hold. This fact is not surprising
because, as stated above, such bounds were not proved rigorously.

The last example is given by $K=3$, $A_{0}=0$, $A_{2}=3/2$, $A_{4}=-5/2$, $%
A_{6}=1/4$, $A_{8}=-1/2$ and $A_{10}=1/4$. In this case there is an exact
ground state $\psi _{0}$ with $E_{0}=0$. The RRVM eigenvalues are shown in
tables \ref{tab:model3e} and \ref{tab:model3o} together with those of Batael
and Drigo Filho.

It is worth comparing the performances of the RRVM and the SSQMGS.
The former approach provides eigenvalues of unlimited accuracy
(depending only on hardware and software facilities) that converge
towards the exact energies from above. On the other hand, the
accuracy of the SSQMGS eigenvalues is determined by the accuracy
of the initial ansatz $\psi _{0}$. Batael and Drigo Filho chose a
particular class of potentials for which one can obtain the exact
$\psi _{0}$ or at least a sufficiently accurate trial function
with the appropriate asymptotic behaviour. Such models are of the
form illustrated in section~\ref {sec:models_methods}. Batael and
Drigo Filho reported more digits than the actual accuracy of their
results. Present RRVM eigenvalues are even more accurate than
those used by Batael and Drigo Filho as benchmark. While it has
already been proved that the RRVM provides upper bounds to the
energies of all the states\cite{M33,F20} such proof is lacking in
the case of the SSQMGS and we have already pointed out two cases
in which the latter approach fails to provide such bounds. As is
well known, one counterexample is sufficient to prove an statement
false.

\begin{table}[tbp]
\caption{RRVM even-state eigenvalues for $A_0=1$, $A_2= A_4=-2$, $A_6=1$
with $\omega =4$}
\label{tab:model1e}
\begin{center}
\par
\begin{tabular}{rD{.}{.}{9}D{.}{.}{9}D{.}{.}{9}D{.}{.}{9}}
\hline \multicolumn{1}{c}{$N$}&\multicolumn{1}{c}{$E_0$}&
\multicolumn{1}{c}{$E_2$} &
\multicolumn{1}{c}{$E_4$} & \multicolumn{1}{c}{$E_6$} \\
\hline
 5&    0.02            &  4.677918651 &  14.53469054 &  28.3757404    \\
10&  6.6\times 10^{-6} &  4.62986462  &  14.35154075 &  27.52416887   \\
15&  1.5\times 10^{-8} &  4.629826578 &  14.3509522  &  27.51712162   \\
20&  8.6\times 10^{-11}&  4.629826494 &  14.35095078 &  27.51709995   \\
25&  9.7\times 10^{-13}&  4.629826493 &  14.35095078 &  27.5170999    \\
30&  2.5\times 10^{-15}&  4.629826493 &  14.35095078 &  27.5170999    \\
$E_n/2$&&                   2.314913246 &  7.17547539  &  13.75854995 \\
Ref.\cite{BF23}&0                 &    2.31799     &  7.18145 &
13.7670

\end{tabular}
\par
\end{center}
\end{table}

\begin{table}[tbp]
\caption{RRVM odd-state eigenvalues for $A_0=1$, $A_2= A_4=-2$, $A_6=1$ with
$\omega =4$}
\label{tab:model1o}
\begin{center}
\begin{tabular}{rD{.}{.}{11}D{.}{.}{10}D{.}{.}{9}D{.}{.}{9}}
\hline \multicolumn{1}{c}{$N$}&\multicolumn{1}{c}{$E_1$}&
\multicolumn{1}{c}{$E_3$} &
\multicolumn{1}{c}{$E_5$} & \multicolumn{1}{c}{$E_7$} \\
\hline
  5 & 0.8655650394 & 9.111949632 & 20.98289274 & 36.23196314    \\
 10 & 0.8459004855 & 9.007614525 & 20.55620684 & 35.17488491    \\
 15 & 0.8458893236 & 9.007557826 & 20.55577168 & 35.16841201    \\
 20 &  0.845889291 & 9.007557632 & 20.55577029 & 35.16839427    \\
 25 & 0.8458892907 & 9.00755763  & 20.55577028 & 35.16839416    \\
 30 & 0.8458892907 & 9.00755763  & 20.55577028 & 35.16839416    \\
 $E_n/2$&0.4229446453 & 4.503778815 & 10.27788514 & 17.58419708 \\
Ref.\cite{BF23}& 0.42388      & 4.50813     & 10.2852     &
17.5941
\end{tabular}
\par
\end{center}
\end{table}

\begin{table}[tbp]
\caption{RRVM even-state eigenvalues for $A_0=0$, $A_2=-26$, $A_4=6$, $A_6=1$
with $\omega=5$}
\label{tab:model2e}
\begin{center}
\par
\begin{tabular}{rD{.}{.}{9}D{.}{.}{9}D{.}{.}{9}D{.}{.}{9}}
\hline \multicolumn{1}{c}{$N$}&\multicolumn{1}{c}{$E_0$}&
\multicolumn{1}{c}{$E_2$} &
\multicolumn{1}{c}{$E_4$} & \multicolumn{1}{c}{$E_6$} \\
\hline
 5 &  -14.39416156 & -2.418081882 & 6.897731829 & 23.83165889   \\
10 &  -14.47163202 & -2.523730405 & 6.599680377 & 21.61724028   \\
15 &  -14.47165595 & -2.523911539 & 6.59851881  & 21.60602543   \\
20 &  -14.47165597 & -2.523911704 & 6.598517525 & 21.60600654   \\
25 &  -14.47165597 & -2.523911705 & 6.598517524 & 21.60600652   \\
30 &  -14.47165597 & -2.523911705 & 6.598517524 & 21.60600652   \\
Ref.\cite{BF23}.& -14.4483     & -2.42763     & 6.596869    &
21.56765

\end{tabular}
\par
\end{center}
\end{table}

\begin{table}[tbp]
\caption{RRVM odd-state eigenvalues for $A_0=0$, $A_2=-26$, $A_4=6$, $A_6=1$
with $\omega =5$}
\label{tab:model2o}
\begin{center}
\begin{tabular}{rD{.}{.}{11}D{.}{.}{10}D{.}{.}{9}D{.}{.}{9}}
\hline \multicolumn{1}{c}{$N$}&\multicolumn{1}{c}{$E_1$}&
\multicolumn{1}{c}{$E_3$} &
\multicolumn{1}{c}{$E_5$} & \multicolumn{1}{c}{$E_7$} \\
\hline
  5 &  -14.3640557 & -0.4515691057 & 13.89792265 & 32.55127969    \\
 10 & -14.42792517 & -0.6900912613 & 13.35318022 & 30.73875482    \\
 15 & -14.42794579 & -0.690175821  & 13.3524621  & 30.7269972     \\
 20 & -14.42794583 & -0.6901759943 & 13.3524612  & 30.72698225    \\
 25 & -14.42794583 & -0.6901759952 & 13.35246119 & 30.72698222    \\
 30 & -14.42794583 & -0.6901759952 & 13.35246119 & 30.72698222    \\
Ref.\cite{BF23}.& -14.4135     & -0.65821      & 13.36402

\end{tabular}
\par
\end{center}
\end{table}

\begin{table}[tbp]
\caption{RRVM even-state eigenvalues for $A_0=0$, $A_2=3/2$, $A_4=-5/2$, $%
A_6=1/4$, $A_8=-1/2$, $A_{10}=1/4$ with $\omega =5$}
\label{tab:model3e}
\begin{center}
\begin{tabular}{rD{.}{.}{11}D{.}{.}{10}D{.}{.}{9}D{.}{.}{9}}
\hline \multicolumn{1}{c}{$N$}&\multicolumn{1}{c}{$E_0$}&
\multicolumn{1}{c}{$E_2$} &
\multicolumn{1}{c}{$E_4$} & \multicolumn{1}{c}{$E_6$} \\
\hline
   5 &   0.09             & 4.573017185 & 16.36066839 & 34.15352004   \\
  10 &  0.002             & 4.32310851  & 15.61645666 & 31.68651075   \\
  15 & 6.1\times 10^{-5}  & 4.315907553 & 15.58461237 & 31.54805825   \\
  20 & 1.8\times 10^{-6}  & 4.315700166 & 15.58363087 & 31.54320834   \\
  25 & 2.1\times 10^-{7}  & 4.31569472  & 15.58360629 & 31.54308125   \\
  30 & 7.6\times 10^{-10} & 4.315694041 & 15.58360331 & 31.54306785   \\
  35 & 1.0\times 10^{-9}  & 4.315694019 & 15.58360321 & 31.54306732   \\
  40 & 1.1\times 10^{-10} & 4.315694016 & 15.58360319 & 31.54306723   \\
  45 & 8.5\times 10^{-12} & 4.315694015 & 15.58360319 & 31.54306722   \\
  50 & 7.6\times 10^{-13} & 4.315694015 & 15.58360319 & 31.54306722   \\
Ref.\cite{BF23}. & 0                  & 4.31612     & 15.5851
& 31.5460
\end{tabular}
\par
\end{center}
\end{table}

\begin{table}[tbp]
\caption{RRVM odd-state eigenvalues for $A_0=0$, $A_2=3/2$, $A_4=-5/2$, $%
A_6=1/4$, $A_8=-1/2$, $A_{10}=1/4$ with $\omega =5$}
\label{tab:model3o}
\begin{center}
\begin{tabular}{rD{.}{.}{11}D{.}{.}{10}D{.}{.}{9}D{.}{.}{9}}
\hline \multicolumn{1}{c}{$N$}&\multicolumn{1}{c}{$E_1$}&
\multicolumn{1}{c}{$E_3$} &
\multicolumn{1}{c}{$E_5$} & \multicolumn{1}{c}{$E_7$} \\
\hline

  5 & 1.256573678 & 9.855150686 & 24.44520554 & 44.68121271   \\
 10 & 1.048870482 & 9.357073321 & 23.02789536 & 41.29594435   \\
 15 & 1.046988529 & 9.351398959 & 23.00064258 & 41.16116412   \\
 20 & 1.046927491 & 9.351217587 & 22.9997951  & 41.15687172   \\
 25 & 1.046922323 & 9.351202299 & 22.99972988 & 41.15661593   \\
 30 & 1.046922115 & 9.351201593 & 22.99972602 & 41.15659472   \\
 35 & 1.046922092 & 9.351201522 & 22.99972569 & 41.15659332   \\
 40 & 1.046922091 & 9.351201519 & 22.99972568 & 41.15659324   \\
 45 & 1.046922091 & 9.351201519 & 22.99972568 & 41.15659323   \\
 50 & 1.046922091 & 9.351201519 & 22.99972568 & 41.15659323   \\
Ref.\cite{BF23}.& 1.04703     &             &             &
\end{tabular}
\par
\end{center}
\end{table}

\end{document}